\documentclass[onecolumn,prd,superscriptaddress,showpacs,floatfix,%
preprintnumbers,nofootinbib,showkeys]{revtex4}

\usepackage{amsmath,amsfonts}
\usepackage{graphicx,subfigure}
\usepackage{epsfig,epsf}
\usepackage{amssymb,color}

\def\nn{\nonumber}

\def\issue(#1,#2,#3){{\bf #1}, #2 (#3)} 

\def\APP(#1,#2,#3){{\rm Acta Phys.\ Polon.} \ \issue({\bf #1},#2,#3)}
\def\ANP(#1,#2,#3){{\rm Annals of Physics} \ \issue({\bf #1},#2,#3)}
\def\ARNPS(#1,#2,#3){{\rm Ann.\ Rev.\ Nucl.\ Part.\ Sci.} \ \issue({\bf #1},#2,#3)}
\def\CPC(#1,#2,#3){{\rm Comp.\ Phys.\ Comm.} \ \issue({\bf #1},#2,#3)}
\def\CIP(#1,#2,#3){{\rm Comput.\ Phys.} \ \issue({\bf #1},#2,#3)}
\def\EPJ(#1,#2,#3){{\rm Eur.\ Phys.\ J.} \ \issue({\bf #1},#2,#3)}
\def\EPJD(#1,#2,#3){Eur.\ Phys.\ J. Direct\ C \ \issue({\bf #1},#2,#3)}
\def\IJMP(#1,#2,#3){{\rm Int.\ J.\ Mod.\ Phys.} \ \issue({\bf #1},#2,#3)}
\def\JHEP(#1,#2,#3){{\rm J.\ High Energy Physics} \ \issue({\bf #1},#2,#3)}
\def\JP(#1,#2,#3){{ J.\ Phys.} \ \issue({\bf #1},#2,#3)}
\def\MPL(#1,#2,#3){{Mod.\ Phys.\ Lett.} \ \issue({\bf #1},#2,#3)}
\def\NP(#1,#2,#3){{Nucl.\ Phys.} \ \issue({\bf #1},#2,#3)}
\def\NIM(#1,#2,#3){{ Nucl.\ Instrum.\ Meth.} \ \issue({\bf #1},#2,#3)}
\def\PL(#1,#2,#3){{ Phys.\ Lett.} \ \issue({\bf #1},#2,#3)}
\def\PR(#1,#2,#3){{ Phys.\ Rev.} \ \issue({\bf #1},#2,#3)}
\def\PRL(#1,#2,#3){{ Phys.\ Rev.\ Lett.} \ \issue({\bf #1},#2,#3)}
\def\SJNP(#1,#2,#3){{ Sov.\ J. Nucl.\ Phys.} \ \issue({\bf #1},#2,#3)}
\def\ZP(#1,#2,#3){{Zeit.\ Phys.} \ \issue({\bf #1},#2,#3)}
\def\be {\begin{equation}}
\def\ee {\end{equation}}
\def\bea {\begin{eqnarray}}
\def\eea {\end{eqnarray}}

\def\bbbar {B-\overline{B}}

\def\bsbsbar {B_s-\overline{B}_s}

\def\bra#1{\left\langle #1\right|}
\def\ket#1{\left| #1\right\rangle}
\newcommand{\bers}{\begin{eqnarray*}}
\newcommand{\eers}{\end{eqnarray*}}
\newcommand{\bt}{\begin{itemize}}
\newcommand{\et}{\end{itemize}}
\def\sla#1{\raise.15ex\hbox{$/$}\kern-.57em #1}

\def\sss{\scriptscriptstyle}

\def\phiM{\phi_{\sss M}}
\def\abar{{\bar a}}
\def\bbar{{\bar b}}
\def\cbar{{\bar c}}

\def\bs{B_s^0}
\def\bsbar{{\overline{B_s^0}}}
\def\barp{{\raise.35ex\hbox
{${\sss (}$}}---{\raise.35ex\hbox{${\sss )}$}}}
\def\bdbarp{\hbox{$B_d$\kern-1.4em\raise1.4ex\hbox{\barp}}}
\def\bsbarp{\hbox{$B_s$\kern-1.4em\raise1.4ex\hbox{\barp}}}

\def\roughly#1{\mathrel{\raise.3ex\hbox
{$#1$\kern-.75em\lower1ex\hbox{$\sim$}}}}

\def\newprd#1#2#3{{ Phys.\ Rev.} {\bf D#1}, #3 (#2)}
\def\m12np{M_{12}^{\rm NP}}
\def\g12np{\Gamma_{12}^{\rm NP}}

\definecolor{purple}{rgb}{0.63,0.13,0.94}
\definecolor{red}{rgb}{1.0,0.0,0.0}
\definecolor{green}{rgb}{0.0,1.0,0.0}
\definecolor{blue}{rgb}{0.0,0.0,1.0}

\begin{document}

\title{
CPT violation and triple-product correlations in B decays
}

\author{Sunando Kumar Patra}

\author{Anirban Kundu}
\affiliation{Department of Physics, University of Calcutta,\\
92, Acharya Prafulla Chandra Road, Kolkata 700009, India. }

\date{\today}

\begin{abstract}

The T-odd triple product (TP) asymmetries in B decays to a pair of vector mesons are treated as a good
probe of CP violation because of the CPT symmetry. If CPT is no longer a good symmetry, such correlations
between T-odd and CP-odd observables do not exist, and one might get unexpected nonzero TP asymmetries
as a signal for CPT violation.  We give a general formalism of TP asymmetries in the presence of CPT violation,
either in decay or in neutral meson mixing. We also
discuss how the observables depending on the transversity amplitudes are modified, and compare our
expressions with the LHCb results, showing that the study of TP asymmetries might turn out to be one of the
best probes for CPT violation.

\end{abstract}

\pacs{11.30.Er, 14.40.Nd}
\keywords{CPT violation, Triple product asymmetry, B decays, Transversity amplitudes}
\maketitle

\section{Introduction}

Triple product (TP) correlations are known to be a good probe of CP violation in $B$ decays
\cite{Kayser,Valencia,bensalem,datta,gronau}. Consider a $B$ meson decaying into two vector
mesons $V_1$ and $V_2$:
\be
B(p) \to V_1(k_1,\varepsilon_1) + V_2(k_2,\varepsilon_2)\,,
\ee
where $k$ and $\epsilon$ are respectively the four-momentum and polarization of the vector mesons.
Suppose one constructs an observable $\alpha \equiv \vec{k}_1.(\vec{\varepsilon}_1\times
\vec{\varepsilon}_2)$, where we have taken out the spatial components of the respective four-vectors.
The asymmetry
\be
\frac{\Gamma(\alpha > 0) - \Gamma(\alpha < 0)}
{ \Gamma(\alpha > 0) + \Gamma(\alpha < 0)}
\ee
is odd under the time-reversal operator T as $\alpha$ itself is T-odd. As CPT is supposed
to be a good symmetry of the Hamiltonian, the asymmetry is CP-odd too, and can be taken as a
probe and measure of CP violation.

TP asymmetries are also an excellent probe of new physics (NP) beyond the Standard Model (SM).
There are many TP asymmetries which are either zero or tiny in the SM but can go up to
observable range under some new physics (NP) dynamics. Also, true TP asymmetries, unlike direct CP
asymmetries, are nonzero even if the strong
phase difference between two competing amplitudes is small or even zero. Of course, TP
asymmetries can be faked by a sizable strong phase difference. The authors of Ref.\ \cite{datta}
have discussed in detail the conditions for observation of TP asymmetries, and also the feasibility
of measuring such asymmetries for different decay channels. The analysis has been extended by the
authors of Ref.\ \cite{gronau} for 4-body final states.

A crucial ingredient of extracting CP-violating signals from TP asymmetries is the CPT theorem:
the combined discrete symmetry CPT, taken in any order, is an exact symmetry of
any local axiomatic quantum field theory (QFT) \cite{streater}.
Experiments have put stringent limits on CPT violation (CPTV), as all tests performed so far
to probe CPTV \cite{pdg} yielded null results \cite{kostel001}.
Still, one should try to measure CPTV in $B$ systems in as many ways as possible,
irrespective of the theoretical dogma, as CPTV can be a flavor-dependent phenomenon, and
the constraints obtained from the
$K$ system \cite{nussinov} may not be applicable to the $B$ systems.
One might also want to know whether any tension between data and the SM expectation is due to
CPT conserving canonical NP, or or just due to CPTV.

%

The issue of CPTV has started to receive significant attention due to the growing phenomenological importance
of CPTV scenarios in neutrino physics and cosmology \cite{others}.
A comprehensive study of CPTV in the neutral $K$ meson system, with a formulation
that is closely analogous to that in the $B$ system, may be found in Ref.~\cite{lavoura}.
CPTV in the $B$ systems, and its possible signatures, including differentiation from CPT conserving
NP models, have been already investigated
by several authors \cite{quark,paschos,patra,patra2}. It was shown that the
lifetime difference of
the two mass eigenstates, or the direct CP asymmetries and semileptonic observables,
may be affected by such new physics. The experimental limits are set by both BaBar, who looked
for diurnal variations of CP-violating observables \cite{cpt-babar}, and Belle, who looked for
lifetime difference of $B_d$ mass eigenstates \cite{cpt-belle}. This makes it worthwhile
to look for possible CPTV effects in the $B_s$ system (by $B_s$ we generically mean both
$\bs$ and $\bsbar$ mesons).

In this paper, we would like to develop the formalism of TP asymmetries with possible CPTV terms in
the Lagrangian. Thus, T violation and CP violation are no longer correlated.
We will show, in detail, how and where deviations occur from the standard CPT
conserving cases. In particular, it will be shown that some decay channels where TP asymmetries are not
expected might throw up new surprises. We will also relate the TP
violating observables with the transversity amplitudes
\cite{datta}, and discuss the implications of the LHCb results \cite{lhcb-cpt} on $B_s\to\phi\phi$.

At this point, we note that violations of different conservation rules lead to different signals. 
For example, violation of $\Delta B = \Delta Q$ keeping CPT invariant would lead to some interesting 
time-integrated dilepton asymmetries \cite{dass}. While a systematic study of the inverse problem,
(i.e. going from the signal to the underlying model) in the $B$ sector is worthwhile, it is outside
the ambit of this paper. We would like to refer the reader to \cite{patra2} for ways to differentiate 
between CPT conserving and CPT violating NP under certain conditions; such a differentiation is 
not always possible.

The paper is arranged as follows. In Section II, we discuss the essential formalism of TP asymmetries when
CPTV terms are present in the decay amplitudes. 
In Section III, we show how the transversity amplitudes are modified by the CPTV terms.
Section IV is devoted to the case where CPTV terms are present in the neutral $B$ meson mixing
Hamiltonian but not in the subsequent decay processes. In Section V, we correlate the expressions with
the data from LHCb. In Section VI, we summarize and conclude.
Some calculational details and a compendium of relevant expressions,
not strictly necessary to catch the main flow of the paper, have been relegated
to the two appendices.

\section{Formalism}

Following Ref.~\cite{datta}, we can write the decay amplitude for $B(p) \to
V_1(k_1,\varepsilon_1) + V_2(k_2,\varepsilon_2)$ as
\be
M = {\bf a} {\cal S} + {\bf b} {\cal D} + i{\bf c} {\cal P}
= {\bf a} \, \varepsilon_1^* \cdot \varepsilon_2^* + \frac{\bf b}{m_B^2}
(p\cdot \varepsilon_1^*) (p\cdot \varepsilon_2^*) + i \frac{\bf c}{m_B^2}
\epsilon_{\mu\nu\rho\sigma} p^\mu q^\nu \varepsilon_1^{*\rho}
\varepsilon_2^{*\sigma} ~,
\label{abcdefs}
\ee
where $q\equiv k_1 - k_2$. Terms are normalized with a
factor $m_B^2$, so that each of ${\bf a}$, ${\bf b}$ and ${\bf c}$ is expected to
be of the same order of magnitude. The ${\bf a}$, ${\bf b}$ and ${\bf c}$ terms correspond to
combinations of $s$, $d$ and $p$-wave amplitudes for the final state, denoted by ${\cal S}$,
${\cal D}$, and ${\cal P}$ respectively. The
quantities ${\bf a}$, ${\bf b}$ and ${\bf c}$ are complex and will in general contain
both CP-conserving strong phases and CP-violating weak phases.

Similarly, the amplitude for the CP-conjugate process
$\bar{B}(p) \to \bar{V}_1(k_1,\varepsilon_1) +
\bar{V}_2(k_2,\varepsilon_2)$ can be expressed as:
\be
{\overline{M}} = {\bf \bar a} \, \varepsilon_1^* \cdot \varepsilon_2^* +
\frac{\bf \bar b}{m_B^2} (p\cdot \varepsilon_1^*) (p\cdot \varepsilon_2^*)
- i \frac{\bf \bar c}{m_B^2} \epsilon_{\mu\nu\rho\sigma} p^\mu q^\nu
\varepsilon_1^{*\rho} \varepsilon_2^{*\sigma} ~,
\label{abcdefsbar}
\ee
where, considering CPT conservation, ${\bf \bar a}$, ${\bf \bar b}$ and ${\bf \bar c}$
can be obtained from ${\bf a}$, ${\bf b}$
and ${\bf c}$ by changing the sign of the weak phases.

In that case, one can write
\begin{align}
\label{abcexplicit}
{\bf a} = \sum_i a_i e^{i \phi_i^a} e^{i \zeta_i^a} & ~~,~~~~ {\bf \bar a} = \sum_i a_i e^{-i \phi_i^a} e^{i \zeta_i^a} ~,
\\
\nn {\bf b} = \sum_i b_i e^{i \phi_i^b} e^{i \zeta_i^b} & ~~,~~~~
{\bf \bar b} = \sum_i b_i e^{-i \phi_i^b} e^{i \zeta_i^b} ~, \\
\nn {\bf c} = \sum_i c_i e^{i \phi_i^c} e^{i \zeta_i^c} & ~~,~~~~
{\bf \bar c} = \sum_i c_i e^{-i \phi_i^c} e^{i \zeta_i^c}
\end{align}
where $\phi_i^{a,b,c}$ ($\zeta_i^{a,b,c}$) are weak (strong) phases of the respective amplitudes.
The relevant quantities for true T-violating TP asymmetries
are $\left[ {\rm Im}({\bf a} {\bf c}^*) - {\rm Im}({\bf \bar a} {\bf \bar c}^*) \right]$ and
$\left[ {\rm Im}({\bf b} {\bf c}^*) - {\rm Im}({\bf \bar b} {\bf \bar c}^*) \right]$, which we get by
adding T-odd asymmetries in $|M|^2$ and $|\overline{M}|^2$. One can show \cite{datta} that
TPs would be non-zero in $B \rightarrow V_1 V_2$ decays as long as ${\rm Im}({\bf a} {\bf c}^*)$ or
${\rm Im}({\bf b} {\bf c}^*)$ is non-zero. For that, both $B\to V_1$ and $B\to V_2$ channels must be present with
different weak phases, following a naive factorization argument, detailed in Appendix A following Ref.~\cite{datta}.

There are two ways to introduce CPT violation in the formalism, namely,
\begin{enumerate}
\item CPTV in the decay amplitude, and
\item CPTV in the mixing amplitude.
\end{enumerate}
We will discuss the former here and postpone the latter for Section IV. However, note that even if
CPTV is present in the decay amplitudes, one can still have a mixing-induced CPT violation, characterized
by time-dependent TP asymmetries, as discussed below.

\subsection{CPTV in Decay}\label{cptvdecay}

Let us start with the first option, which can be subdivided into two categories:

\subsubsection{Type I: CPTV present only in the $p$-wave amplitude}

We introduce the CPTV parameter $f \equiv {\rm Re}(f) +i {\rm Im}(f)$ in the following way:
\be
{\bf c} = \sum_i c_i e^{i \phi_i^c} e^{i \zeta_i^c} (1 - f)\,,\ \
{\bf \bar c} = \sum_i c_i e^{-i \phi_i^c} e^{i \zeta_i^c} (1 + f^*)\,,
\ee
and other amplitudes remaining the same. This is the simplest way to introduce CPTV; a channel-dependent
CPTV parameter $f_i$ would only complicate the calculation without giving any extra insight.

The relevant quantity for TP is
\begin{align}
\nn \frac12\left[ {\rm Im}({\bf a} {\bf c}^*) - {\rm Im}({\bf \bar a} {\bf \bar c}^*)
\right] &= \sum_{i,j} a_i c_j [\sin \left( \phi_i^a - \phi_j^c
\right) \cos \left( \zeta_i^a - \zeta_j^c \right) - {\rm Re}(f) \cos \left( \phi_i^a - \phi_j^c
\right) \sin \left( \zeta_i^a - \zeta_j^c \right)\\
& - {\rm Im}(f) \sin \left( \phi_i^a - \phi_j^c
\right) \sin \left( \zeta_i^a - \zeta_j^c \right)]\,.
\label{TPform2}
\end{align}
A similar expression is obtained for $\frac12\left[ {\rm Im}({\bf b} {\bf c}^*) - {\rm Im}({\bf \bar b} {\bf \bar c}^*)
\right]$. Even if the weak phase difference vanishes, these are still nonzero because of the second term,
so the TP asymmetry will essentially probe ${\rm Re}(f)$.

\subsubsection{Type II: Universal CPTV present in all amplitudes}

In this case, the coefficients from Eqs.~(\ref{abcdefs}) and (\ref{abcdefsbar}) are modified as
\be
({\bf a}, {\bf b}, {\bf c}) \to ({\bf a}, {\bf b}, {\bf c}) (1-f)\,,\ \
(\bar {\bf a}, \bar {\bf b}, \bar {\bf c}) \to  (\bar {\bf a}, \bar {\bf b}, \bar {\bf c}) (1 + f^\ast)\,.
\ee
Thus, the relevant expression for TP becomes,
\begin{align}
\frac12\left[ {\rm Im}({\bf a} {\bf c}^*) - {\rm Im}({\bf \bar a} {\bf \bar c}^*)
\right] &= \sum_{i,j} a_i c_j [\sin \left( \phi_i^a - \phi_j^c
\right) \cos \left( \zeta_i^a - \zeta_j^c \right) - 2 {\rm Re}(f) \cos \left( \phi_i^a - \phi_j^c
\right) \sin \left( \zeta_i^a - \zeta_j^c \right)]
\label{TPform3}
\end{align}
Here too, only the second term remains in absence of weak phase.

Following Eq.~(\ref{abc}) taken from \cite{datta}, one finds the cases where no TP asymmetry is expected in the SM.
On the other hand, introduction of CPTV may induce nonzero TP asymmetries for some of the cases as follows:
 \begin{enumerate}
\item In order to have a TP correlation in a given decay,
both of the amplitudes in eq.~(\ref{2amps}) must be present, otherwise either
$X$ or $Y$ becomes zero. This remains true for CPTV of type II, but for type I,
even in the absence of either $X$ or $Y$, TPs can be generated.

\item For the same reason as above, CPTV of type I can produce nonzero TPs even if
$V_1$ and $V_2$ have identical flavor wavefunctions (same meson, or an excited state).
Such nonzero TPs are not allowed in the SM as then
${\bf a}$, ${\bf b}$, and ${\bf c}$ are all proportional to the same factor and there is no
relative phase.

\item In the SM (or in any NP model with CPT conservation), two kinematical amplitudes must
have different weak phases for a nonzero TP asymmetry.
Thus, if the quark-level decay is dominated by a single decay amplitude, a nonzero TP can never be generated.
This is again not necessarily true for CPTV of either type I or type II,
as we have seen from Eqs.~(\ref{TPform2})
and (\ref{TPform3}) that even in the absence of weak phase difference, one of the terms
in the relevant expressions can have a nonzero value.

\end{enumerate}

\subsubsection{Effects of Type I and Type II CPTV in mixing}

There could be another way to induce CPTV. Let us suppose CPTV to be present only for
$B \to V_1$ and not for $B \to V_2$. As can be seen from Eq.~(\ref{term1}), this changes only the terms with
the same phase in the expressions for ${\bf a}$, ${\bf b}$, and ${\bf c}$.
Thus, $|f|$ is absorbed in the form factors and $\arg(f)$ in the phase. Obviously, this scenario does not produce any TP
even if CPTV is present.


Now let us consider the special case where $V_1$
can be accessed from $B$ but not from $\bar{B}$, and vice versa. Let us also take, for simplicity, $B\to V_1$ and
$\bar{B}\to V_2$ to be single-amplitude processes. For $B=B_{d,s}$, there will be a mixing-induced
TP because the $B$ meson can oscillate into $\bar{B}$ and hence decay to $V_2$, thus providing the second
amplitude. The relevant T-violating
terms, as shown in Ref.~\cite{datta}, are proportional to the ${\bf a}$-${\bf c}$ (and ${\bf b}$-${\bf c}$) interference
contributions, and are given by
\begin{align}
\label{timedepTP}
|M|^2_{{\bf ac}} + |{\overline M}|^2_{\bf ac} &\sim {\rm Im}({\bf a} {\bf c}^*) - {\rm Im}({\bf \bar a} {\bf \bar c}^*)\nn\\
& = \cos^2 \left(\frac{\Delta M t}{2} \right) {\rm Im}({\bf a}_1
{\bf c}_1^* - {\bf \abar}_1 {\bf \cbar}_1^*)
+ \sin^2 \left(\frac{\Delta M t}{2}\right) {\rm Im}({\bf a}_2 {\bf c}_2^* - {\bf \abar}_2 {\bf \cbar}_2^*) \nn\\
& + \sin \left(\frac{\Delta M t}{2} \right) \cos\left(\frac{\Delta M t}{2} \right)
{\rm Re} \left[e^{-2 i \phiM}
{\bf a}_2 \, {\bf c}_1^* - e^{2 i \phiM} {\bf \abar}_2 \, {\bf \cbar}_1^*
- e^{2 i \phiM} {\bf a}_1 \, {\bf c}_2^* + e^{-2 i \phiM} {\bf \abar}_1 \, {\bf \cbar}_2^*\right]\,.
\end{align}
where $\Delta M$ is the mass difference of the two eigenstates, and following Eq.~(\ref{abcdefs}),
\begin{align}
\label{BVVamps}
A(B \to V_1 V_2) & = {\bf a}_1 {\cal S} + {\bf b}_1 {\cal D} + i {\bf c}_1 {\cal P},~~~~~~~
A(\bar{B} \to V_1 V_2) = {\bf a}_2 {\cal S} + {\bf b}_2 {\cal D} + i {\bf c}_2 {\cal P}, \nn\\
A(B \to {\bar V}_1 {\bar V}_2) & = {\bf \abar}_2 {\cal S} + {\bf \bbar}_2 {\cal D} - i
  {\bf \cbar}_2 {\cal P},~~~~~~~ A(\bar{B} \to {\bar V}_1 {\bar V}_2) = {\bf \abar}_1 {\cal S} + {\bf \bbar}_1 {\cal D} - i
  {\bf \cbar}_1 {\cal P},
\end{align}
so that,
\bea
M \equiv A(B(t) \to V_1 V_2) = e^{-i \left( M - \frac{i}{2}\Gamma
\right) t} \left[ {\bf a} {\cal S} + {\bf b} {\cal D} + i {\bf c} {\cal P} \right], \nn\\
{\bar M} \equiv A(\bar{B}(t) \to {\bar V}_1 {\bar V}_2) = e^{-i
\left( M - \frac{i}{2}\Gamma \right) t} \left[ {\bf \bar a} {\cal S} + {\bf \bar b} {\cal D} - i
{\bf \bar c} {\cal P} \right],
\eea
with
\begin{align}
{\bf a} & = {\bf a}_1 \cos \left({\Delta Mt \over 2} \right) - i \, e^{-2 i
  \phiM} \sin \left({\Delta Mt \over 2} \right) {\bf a}_2 ~,~~~~{\bf \bar a} = {\bf \abar}_1 \cos \left({\Delta Mt \over 2}
\right) - i \,
  e^{2 i \phiM} \sin \left({\Delta Mt \over 2} \right) {\bf \abar}_2 ~, \nn\\
{\bf b} & = {\bf b}_1 \cos \left({\Delta Mt \over 2} \right) - i \, e^{-2 i
  \phiM} \sin \left({\Delta Mt \over 2} \right) {\bf b}_2 ~,~~~~{\bf \bar b} = {\bf \bbar}_1 \cos \left({\Delta Mt \over 2} \right) - i \,
  e^{2 i \phiM} \sin \left({\Delta Mt \over 2} \right) {\bf \bbar}_2 ~, \nn\\
{\bf c} & = {\bf c}_1 \cos \left({\Delta Mt \over 2} \right) - i \, e^{-2 i
  \phiM} \sin \left({\Delta Mt \over 2} \right) {\bf c}_2 ~,~~~~{\bf \bar c} = {\bf \cbar}_1 \cos \left({\Delta Mt \over 2} \right) - i \,
  e^{2 i \phiM} \sin \left({\Delta Mt \over 2} \right) {\bf \cbar}_2\,.
\end{align}
Note that amplitudes like ${\bf a}_1$ are complex, with relevant weak and strong phases:
\be
{\bf a}_1 = {a}_1 e^{i \phi_1^a} e^{i \zeta_1^a}\,,
\ee
The first term in eq.(\ref{timedepTP}) describes the time evolution of the TP in $B \to V_1
V_2$ and the second term, generated due to $B$--$\bar{B}$ mixing, describes the
time evolution of the TP in $\bar{B} \to V_1 V_2$. The third term
can potentially generate a TP due to $B$--$\bar{B}$ mixing even in the absence of TP in $B \to V_1 V_2$.
This term can be rewritten after explicitly writing down ${\bf a}_1$, ${\bf a}_2$ etc.\ following Eq.~(\ref{abcexplicit}):
\begin{align}
\label{mixingTP}
-(\sin \Delta M t) [ a_{2} c_{1} \sin(
  \phi^{a}_2 - \phi^{c}_1 - 2 \phiM) \sin(\zeta^{a}_2 -
  \zeta^{c}_1) - a_{1} c_{2} \sin( \phi^a_1 - \phi^c_2 + 2 \phiM)
  \sin(\zeta^a_1 - \zeta^c_2)].
\end{align}
This expression goes to zero in the absence of strong phase differences,
which is intuitively obvious as strong phase differences are related in part to kinematics,
and the TP vanishes if kinematics of $\bar{B}  \rightarrow V_2$ is identical to $B \rightarrow V_1$.

However, in the presence of CPTV of Type I, the expression in (\ref{mixingTP}) is modified to
\begin{align}
\label{mixingTPCPT1}
\nn & - (\sin \Delta M t) [ a_{2} c_{1} \sin(
  \phi^a_2 - \phi^c_1 - 2 \phiM) \sin(\zeta^a_2 -
  \zeta^c_1) - a_{1} c_{2} \sin( \phi^a_1 - \phi^c_2 + 2 \phiM)
  \sin(\zeta^a_1 - \zeta^c_2)]\\
\nn & ~~~ - 2 {\rm Re}(f) [a_{2} c_{1} \cos(
  \phi^a_2 - \phi^c_1 - 2 \phiM) \cos(\zeta^a_2 -
  \zeta^c_1) - a_{1} c_{2} \cos( \phi^a_1 - \phi^c_2 + 2 \phiM)
  \cos(\zeta^a_1 - \zeta^c_2)]\\
 & ~~~ - {\rm Im}(f) [a_{2} c_{1} \sin(
  \phi^a_2 - \phi^c_1 - 2 \phiM) \cos(\zeta^a_2 -
  \zeta^c_1) - a_{1} c_{2} \sin( \phi^a_1 - \phi^c_2 + 2 \phiM)
  \cos(\zeta^a_1 - \zeta^c_2)]\,,
\end{align}
while for CPTV of Type II, the same expression takes the form
\begin{align}
\label{mixingTPCPT2}
\nn & - (\sin \Delta M t)[ a_{2} c_{1} \sin(
  \phi^a_2 - \phi^c_1 - 2 \phiM) \sin(\zeta^a_2 -
  \zeta^c_1) - a_{1} c_{2} \sin( \phi^a_1 - \phi^c_2 + 2 \phiM)
  \sin(\zeta^a_1 - \zeta^c_2)]\\
& ~~~ - 2 {\rm Re}(f) [a_{2} c_{1} \cos(
  \phi^a_2 - \phi^c_1 - 2 \phiM) \cos(\zeta^a_2 -
  \zeta^c_1) - a_{1} c_{2} \cos( \phi^a_1 - \phi^c_2 + 2 \phiM)
  \cos(\zeta^a_1 - \zeta^c_2)]\,.
\end{align}
The last two equations show that in the presence of CPTV, we can get a non-zero TP
from mixing, even if the strong phase differences vanish. Only if the final state is
self-conjugate, the third term in eq.(\ref{timedepTP}) is zero and the first two terms add up,
so the TP in $B \to V_1 V_2$ is time-independent and this remains true even in the presence of CPTV.

\section{Relation to Transversity Amplitudes}

The angular momentum amplitudes are related to the tranversity amplitudes by the following relations \cite{datta}:
\be
\label{Aidefs}
A_\| = \sqrt{2} {\bf a}\,,~~~ A_0 = -{\bf a} x - {m_1 m_2 \over m_{\sss B}^2} {\bf b}
(x^2 - 1)\,,~~~
A_\perp = 2\sqrt{2} \, {m_1 m_2 \over m_{\sss B}^2} {\bf c}
\sqrt{x^2 - 1}\,.
\ee
Let us consider, following Ref.~\cite{gronau}, the channels in which each of the two vector mesons
in $B\to V_1V_2$ further decays into two pseudoscalar mesons.
The decay angular distribution in three dimensions is given in terms of the three transversity amplitudes.
We take $\theta_1$($\theta_2$) to be the angle between the direction of motion of $P_1$ ($P_2$) in the $V_1$ ($V_2$)
rest frame and that of $V_1$ ($V_2$) in the $B$ rest frame. The angle between the planes defined by $P_1P'_1$
and $P_2P'_2$ in the $B$ rest frame is denoted by $\varphi$.
One obtains \cite{gronau}
\begin{eqnarray}
\label{angular}
\frac{d\Gamma}{d\cos\theta_1d\cos\theta_2 d\varphi} &=&
N\big[|A_0|^2\cos^2\theta_1\cos^2\theta_2 +
\frac{|A_\||^2}{2}\sin^2\theta_1\sin^2\theta_2\cos^2\varphi + \frac{|A_\perp|^2}{2}\sin^2\theta_1\sin^2\theta_2\sin^2\varphi \nn\\
&&+ \frac{{\rm Re}(A_0A^*_\|)}{2\sqrt2}\sin 2\theta_1\sin 2\theta_2\cos\varphi - \frac{{\rm Im}(A_\perp A^*_0)}{2\sqrt2}\sin 2\theta_1\sin 2\theta_2\sin\varphi \nn\\
&& -  \frac{{\rm Im}(A_\perp A^*_\parallel)}{2}\sin^2\theta_1 \sin^2\theta_2\sin 2\varphi \big]\,,\nn\\
 \frac{d\bar \Gamma}{d\cos\bar\theta_1d\cos\bar\theta_2 d\bar\varphi} &=&
N\big[|\bar A_0|^2\cos^2\bar\theta_1\cos^2\bar\theta_2 +
\frac{|\bar A_\perp|^2}{2}\sin^2\bar\theta_1\sin^2\bar\theta_2\sin^2\bar\varphi + \frac{|\bar A_\parallel|^2}{2}\sin^2\bar\theta_1\sin^2\bar\theta_2 \cos^2\bar\varphi  \nn\\
&&+\frac{{\rm Re}(\bar A_0\bar A^*_\parallel)}{2\sqrt2}
\sin 2\bar\theta_1\sin 2\bar\theta_2\cos\bar\varphi + \frac{{\rm Im}(\bar A_\perp \bar A^*_0)}{2\sqrt2}\sin 2\bar \theta_1 \sin 2\bar \theta_2\sin\bar\varphi \nn\\
&& +  \frac{{\rm Im}(\bar A_\perp \bar
A^*_\parallel)}{2}\sin^2\bar \theta_1\sin^2\bar\theta_2\sin 2\bar\varphi \big]\,.
\end{eqnarray}

Integrating these over $\theta_1$ and $\theta_2$ gives a T-odd asymmetry involving
$\sin 2\varphi$ \cite{datta}
\begin{align}
\label{A2_T}
A^{(2)}_T \equiv \frac{\Gamma(\sin 2\varphi >0) - \Gamma(\sin 2\varphi<0)}
{\Gamma(\sin 2\varphi >0) + \Gamma(\sin 2\varphi<0)} = -\frac{4}{\pi}\frac{{\rm Im}(A_\perp A^*_\parallel)}{|A_0|^2
+ |A_\perp|^2 + |A_\parallel|^2}~.
\end{align}

Similarly, we may define an asymmetry with respect to the values of $\sin\varphi$, assigning it
the sign of $\cos\theta_1\cos\theta_2$ and integrating over all angles,
\begin{align}
\label{A1_T}
A^{(1)}_T &\equiv \frac{\Gamma[{\rm sign}(\cos\theta_1\cos\theta_2)\sin\varphi > 0]
- \Gamma[{\rm sign}(\cos\theta_1\cos\theta_2)\sin\varphi < 0]}
{\Gamma[{\rm sign}(\cos\theta_1\cos\theta_2)\sin\varphi > 0]
+ \Gamma[{\rm sign}(\cos\theta_1\cos\theta_2)\sin\varphi < 0]} \nn\\
&= -\frac{2\sqrt2}{\pi}\frac{{\rm Im}(A_\perp A^*_0)}{|A_0|^2 +
 |A_\perp|^2 + |A_\parallel|^2}\,.
\end{align}
One can define similar asymmetries $\bar A^{(1)}_T$ and $\bar A^{(2)}_T$ by integrating the second part of
Eq.~(\ref{angular}) and proceeding in a similar manner. As the $p$-wave amplitude in ${\overline{M}}$ changes sign
relative to that of $M$ (Eqs.~(\ref{abcdefs}) and (\ref{abcdefsbar})), the sign of the
T-odd asymmetry in $|{\overline{M}}|^2$ is opposite that in
$|M|^2$. The true T-violating asymmetry is therefore found by {\it
adding} the T-odd asymmetries in $|M|^2$ and $|{\overline{M}}|^2$
\cite{Valencia}:
\be
{\cal A}_{\sss T} \equiv \frac12 \left (A_T + {\bar A}_T\right)\,.
\label{Tviolasym}
\ee
This essentially means that instead of ${\rm Im}(A_\perp A^*_i)$, we should look for expressions involving ${\rm Im}(A_\perp A^*_i + \bar A_\perp\bar A^*_i)$ in search of true TP-violating asymmetries. If we consider specifically the decay $B_s \to \phi\phi$, following Ref.~\cite{gronau}, we notice that final states are flavorless and
accessible to both $B_s$ and $\bar B_s$. As a result of $B_s$--$\bar B_s$ oscillation, the angular
decay distributions become time-dependent. Using standard notations for $B_s$--$\bar B_s$ mixing,
and assuming no CP violation in mixing ($|q/p|=1$) and decay ($|\bar A_k|=|A_k|$), one
has~\cite{Branco:1999fs}
\be\label{phik}
\frac{q}{p}\frac{\bar A_k}{A_k} = \eta_k e^{-2i\phi_k}\,.
\ee
Here $\eta_k$ is the CP parity for a state of transversity $k$ ($\eta_0 =
\eta_\parallel= -\eta_\perp =+1$), while $\phi_k$ is the weak phase involved in
an interference between mixing and decay amplitudes. Denoting the CP conserving
strong phase of $A_k$ by $\zeta_k$, one can write $A_k=|A_k|e^{i\zeta_k}e^{i\phi_k}$,
so that $\bar A_k = (p/q) \eta_k e^{i \zeta_k} e^{-i \phi_k}$.
One thus has for $i=0, \parallel$:
\be
\label{timeind}
{\rm Im}(A_\perp A^*_i + \bar A_\perp\bar A^*_i) = |A_\perp| |A_i|{\rm Im} \left [
e^{i\zeta^-} (e^{i\phi^-}
- e^{-i\phi^-})\right]
= 2|A_\perp| |A_i|\cos(\zeta^-) \sin(\phi^-)\,,
\ee
where we define the notations for our future references:
\be
\zeta^- = \zeta_\perp - \zeta_i\,, \ \ \
\phi^- = \phi_\perp - \phi_i\,,\ \ \
\phi^+ = \phi_\perp + \phi_i\,.
\label{shorthand}
\ee
One finds from Eq.~(\ref{Aidefs}) that expressions such as ${\rm Im}(A_\perp A^*_0)$ are proportional to linear
combinations of terms like ${\rm Im}({\bf a^*} {\bf c})$ and ${\rm Im}({\bf b^*} {\bf c})$. Now, as per Eq.~(\ref{abc}),
they are all zero for decays like $B_s\to\phi\phi$; thus, $A^{(1)}_T$, $A^{(2)}_T$,  and consequently all of their
combinations are zero. This can also be seen from Eq.~(\ref{timeind}) if the weak phases for all the
tranversity amplitudes are the same. So, any nonzero values to any of these observables unambiguously point to new
physics.

Let us assume the NP to be CPT violating in nature, and parametrize the amplitudes following
Eqs.~(\ref{abcexplicit}) and  (\ref{Aidefs}):

\begin{eqnarray}
\label{Apexp}
&& A_\perp = \sum_l \left|A_\perp^l\right| e^{i \phi_\perp^l} e^{i \zeta_\perp^l} (1 - f)\,,
\ \ \ \ \ A_i = \sum_m |A_i^m| e^{i\zeta_i^m} e^{i\phi_i^m},\nn\\
&& \bar A_\perp = \eta_\perp \sum_l \left|A_\perp^l\right| e^{- i \phi_\perp^l} e^{i \zeta_\perp^l} (1 + f^*)\,,
\ \ \ \ \ \bar A_i = \sum_m |A_i^m| e^{i\zeta_i^m} e^{-i\phi_i^m}\,,\ \ \ \
(i=0,\parallel)\,.
\end{eqnarray}
Using the notation $\zeta_{l,m}^- = \left(\zeta_\perp^l - \zeta_i^m\right)$ and $\phi_{l,m}^- =
 \left(\phi_\perp^l - \phi_i^m\right)$, we obtain,
\begin{align}
\label{cptcase}
\nn {\rm Im}(A_\perp A^*_{i} + \bar A_\perp \bar A^*_{i}) &= 2~\sum_{l,m}
|A_\perp^l| |A_{i}^m| \big[\sin(\phi_{l,m}^-) \cos(\zeta_{l,m}^-) -
{\rm Re}(f) \sin(\zeta_{l,m}^-) \cos(\phi_{l,m}^-) \\
&+ {\rm Im}(f) \sin(\phi_{l,m}^-) \sin(\zeta_{l,m}^-)\big]\,.
\end{align}
For $f=0$ this reduces to Eq.~(\ref{timeind}). On the other hand, even if $\phi_{l,m}^- = 0$,
we still get a nonzero result:
\be
{\rm Im}(A_\perp A^*_{i} + \bar A_\perp \bar A^*_{i}) =
-2~\sum_{l,m} |A_\perp^l| |A_{i}^m| {\rm Re}(f) \sin(\zeta_{l,m}^-)\,.
\ee

\subsection{Time dependence of the Transversity Amplitudes}

Next, let us consider the time dependence of transversity amplitudes; we will use a formalism closely following
Ref.~\cite{gronau}. The states $B$ and $\bar{B}$ evolve in time as
\be
B(t) = f_+(t)B + (q/p)f_-(t)\bar B~,~~~~~~~\bar B(t) = (p/q)f_-(t)B + f_+(t)\bar B\,,
\ee
where
\begin{align}
\label{g^2}
\nn f_+(t) &= \frac{1}{2} \left(e^{- i \lambda^{(q)}_1 t} + e^{- i \lambda^{(q)}_2 t}\right) = 
\frac{1}{2} \left(e^{- i m_1 t - (\Gamma_1 t / 2)} + e^{- i m_2 t - (\Gamma_2 t / 2)}\right)\\
\nn f_-(t) &= \frac{1}{2} \left(e^{- i \lambda^{(q)}_1 t} - e^{- i \lambda^{(q)}_2 t}\right) = 
\frac{1}{2} \left(e^{- i m_1 t - (\Gamma_1 t / 2)} - e^{- i m_2 t - (\Gamma_2 t / 2)}\right),\\
|f_\pm(t)|^2 &= (e^{-\Gamma t}/2)[\cosh(\Delta\Gamma t/2) \pm \cos(\Delta Mt)],
\nonumber\\
f^*_+(t)f_-(t) &= (e^{-\Gamma t}/2)[\sinh(\Delta\Gamma t/2) - i\sin(\Delta M t)]\,,
\end{align}
$\Delta M$ and $\Delta\Gamma$ being the mass and width differences of the stationary states respectively.

Time dependence of transversity amplitudes, $A_k\equiv \langle k|B\rangle,
\bar A_k\equiv \langle k|\bar B\rangle$~($k=0, \parallel, \perp$), is given by:
\bea\label{amptime}
A_k(t) &\equiv & \langle k | B(t) \rangle = f_+(t) A_k +(q/p) f_-(t) \bar A_k\,,
\nonumber\\
\bar A_k(t) & \equiv & \langle k | \bar B(t) \rangle = (p/q)f_-(t)A_k + f_+(t)\bar A_k\,.
\eea

Let us calculate the interference terms $A^*_i(t)A_k(t)$ and $ \bar A^*_i(t)\bar A_k(t)$, where
$i=0, \parallel, k=\perp$. Inserting
$A^*_iA_k = |A_i||A_k| (1 - f) \exp[{i(\zeta_k-\zeta_i)}]\exp[i(\phi_k-\phi_i)]$\,, and
$\bar A^*_i\bar A_k =  \eta_i\eta_k|A_i||A_k| (1 + f^*)
\exp[i(\zeta_k-\zeta_i)]\exp[-i(\phi_k-\phi_i)]\,,$
one gets, using Eq.~(\ref{shorthand}),
\begin{align}
\label{untagged}
\nn {\rm Im}[A_\perp(t) A^*_i(t) + \bar A_\perp(t) \bar A^*_i(t)] &=  2|A_\perp||A_i|e^{-\Gamma t} \times \\
\nn &\big[\left\{\cos(\zeta^-) \sin(\phi^-) - \sin(\zeta^-) \left({\rm Re}(f) \cos(\phi^-) - {\rm Im}(f) \sin(\phi^-)
\right)\right\}\cosh(\Delta\Gamma t/2) \\
 &+ \left\{\cos(\zeta^-) \sin(\phi^+)+ \sin(\zeta^-) \left({\rm Re}(f) \cos(\phi^+) + {\rm Im}(f) \sin(\phi^+)
\right)\right\}\sinh(\Delta\Gamma t/2)\big]\,.
\end{align}

This, again, agrees with Eq.~(\ref{timeind}) at $t=0, f=0$. When CPT is conserved, it shows the
variation of a genuine CP violating quantity with time which requires no strong phase differences.
The CPTV contribution is nonzero even if the weak phase difference vanishes but the strong phase
difference $\zeta^-$ must be nonzero.

If there are more than one decay channel contributing to the transversity amplitudes, Eq.~(\ref{untagged})
can be generalized to
\begin{align}
\label{untagdirect}
\nn {\rm Im}[A_\perp(t) A^*_i(t) + \bar A_\perp(t) \bar A^*_i(t)] &= \sum_{l,m} 2|A^l_\perp||A^m_i|e^{-\Gamma t}
\times\\
\nn \big[ & \left\{\cos(\zeta_{l,m}^-) \sin(\phi^-_{l,m}) - \sin(\zeta_{l,m}^-)
\left[{\rm Re}(f) \cos(\phi^-_{l,m}) - {\rm Im}(f) \sin(\phi^-_{l,m})\right]\right\}
\cosh(\Delta\Gamma t/2) \\
 + &\left\{\cos(\zeta_{l,m}^-) \sin(\phi^+_{l,m})+ \sin(\zeta_{l,m}^-) \left[{\rm Re}(f) \cos(\phi^+_{l,m})
+ {\rm Im}(f) \sin(\phi^+_{l,m})\right]\right\}\sinh(\Delta\Gamma t/2)\big]\,.
\end{align}

The two ``true" CP violating time-integrated triple product asymmetries ($i=0, \parallel$) for untagged
decays are proportional to
\begin{align}
\label{int}
\nn \Gamma \int_0^\infty {\rm Im}[A_\perp(t) A^*_i(t) &+ \bar A_\perp(t) \bar A^*_i(t)]dt =
\sum_{l,m} 2|A^l_\perp||A^m_i| \times \\
\nn &\big[\left\{ \cos(\zeta_{l,m}^-) \sin(\phi^-_{l,m}) - \sin(\zeta_{l,m}^-) \left({\rm Re}(f) \cos(\phi^-_{l,m}) -
{\rm Im}(f) \sin(\phi^-_{l,m})\right)\right\} \\
\nn &+\left\{ \cos(\zeta_{l,m}^-) \sin(\phi^+_{l,m})+ \sin(\zeta_{l,m}^-) \left({\rm Re}(f) \cos(\phi^+_{l,m}) +
{\rm Im}(f) \sin(\phi^+_{l,m})\right)\right\}(\Delta\Gamma/2\Gamma) \\
&+ {\cal O}[(\Delta\Gamma/2\Gamma)^2]\big]\,.
\end{align}

In the limit $\Delta\Gamma \ll \Gamma$, one can neglect everything apart from the first term in
Eq.~(\ref{int}) and finds
\begin{align}
\label{adecay}
\nn \mathcal{A}^{(1) {\rm untagged}}_T &= - \frac{4 \sqrt2}{\pi} \sum_{l,m}
\frac{|A^l_\perp||A^m_0|\left[ \cos(\zeta^{0-}_{l,m}) \sin(\phi^{0-}_{l,m}) - \sin(\zeta^{0-}_{l,m})
\left({\rm Re}(f) \cos(\phi^{0-}_{l,m}) - {\rm Im}(f) \sin(\phi^{0-}_{l,m})\right)\right]}{\left(|A_0|^2 +
 |A_\perp|^2 + |A_\parallel|^2\right) + \left(|\bar A_0|^2 +
 |\bar A_\perp|^2 + |\bar A_\parallel|^2\right)}\\
\nn & + {\cal O}[(\Delta\Gamma/2\Gamma)]\\
\mathcal{A}^{(2) {\rm untagged}}_T &= - \frac{8}{\pi} \sum_{l,m}
\frac{|A^l_\perp||A^m_\parallel|\left[ \cos(\zeta^{\parallel -}_{l,m}) \sin(\phi^{\parallel-}_{l,m}) - \sin(\zeta
^{\parallel -}_{l,m}) \left({\rm Re}(f) \cos(\phi^{\parallel-}_{l,m}) - {\rm Im}(f) \sin(\phi^{\parallel-}_{l,m})
\right)\right]}{\left(|A_0|^2 +
 |A_\perp|^2 + |A_\parallel|^2\right) + \left(|\bar A_0|^2 +
 |\bar A_\perp|^2 + |\bar A_\parallel|^2\right)}\\
\nn & + {\cal O}[(\Delta\Gamma/2\Gamma)]\,,
\end{align}
where
$\zeta^{i-}_{l,m}= (\zeta^l_\perp - \zeta^m_i)$ and $\phi^{i-}_{l,m}= (\phi^l_\perp - \phi^m_i)$ for $i= 0, \parallel$,
and the coefficients of the $\Delta\Gamma/2\Gamma$ terms can be easily found out from Eq.~(\ref{int}).

In the absence of weak phase difference, $\phi_\perp =\phi_0 =\phi_\parallel$, {\em i.e.} $\phi^{i-}_{l,m}=0$,
the asymmetries vanish in the leading order if CPT is conserved \cite{gronau} but is nonzero if CPT is violated.
Again, a nonzero strong phase difference $\zeta_{l,m}^{i-}$ is obligatory for this.

In the SM, all the three transversity amplitudes have approximately equal and very small weak phases.
Thus, one expects the asymmetries to be quite small. On the other hand, if CPTV is present, these
asymmetries, measured in self-tagged decays to final CP eigenstates, need not be nonzero; thus,
measurements of such asymmetries may either put stringent limits on the CPT violating parameter $f$,
or indicate physics beyond SM.

\section{CPT Violation in Mixing}

One can also consider the case where CPTV is present not in decay but in $\bbbar$ mixing, and parametrize the $2\times 2$
Hamiltonian matrix with the introduction of an extra complex parameter $\delta$ which incorporates CPT violation
\cite{patra}:
\be
\delta = \frac{H_{22}-H_{11}}{\sqrt{H_{12}H_{21}}}\,,
\ee
so that
\be
{\cal M} = \left[\left(
 \begin{array}{cc}
 M_0-\delta' & M_{12}\\ M_{12}^* & M_0+\delta'
 \end{array} \right)
- \frac{i}{2} \left(
 \begin{array}{cc}
 \Gamma_0 & \Gamma_{12}\\ \Gamma_{12}^* & \Gamma_0
 \end{array} \right)\right]\,,
\ee
where $\delta'$ is defined by
\be
\delta = \frac{2\delta'}{\sqrt{H_{12}H_{21}}}\,.
\ee
We work within the Wigner-Weisskopf approximation which is a reliable one after a time scale 
of $\sim 1/M_B$. Violation of this approximation, which has nevertheless been considered in the 
literature \cite{dass-grimus}, would change all the subsequent expressions, and we refrain from 
considering such a possibility. This will give, akin to the Bell-Steinberger analysis \cite{bell},
a way to measure the CPT violating parameter $\delta$ in terms of the interference amplitudes 
which are supposed to be good probes of CP violation. 

Eq.~(\ref{amptime}) can be written as
\bea
A_i(t) &\equiv & \langle k | B(t) \rangle = f_+(t) A_i + \eta_1 f_-(t) \bar A_i\,,
\nonumber\\
\bar A_i(t) & \equiv & \langle k | \bar B(t) \rangle = \frac{f_-(t)}{\eta_2}A_i + \bar f_+(t)\bar A_i\,,
\eea
where $f_{\pm}(t)$, $\bar f_+(t)$ and $\eta_{(1,2)}$ are defined in Appendix \ref{cptmixform}.
Using Eq.~(\ref{shorthand}), one gets,

\begin{align}
{\rm Im}[A_\perp(t) A^*_i(t) + \bar A_\perp(t) \bar A^*_i(t)] &= 2 e^{-\Gamma t} |A_i||A_\perp|\times \nn\\
&\left[\cosh(\Delta\Gamma t/2) \left\{\cos\zeta^- \sin\phi^- - \frac{1}{4}{\rm Im}\delta \cos\phi^+
\left(1 + \sin\zeta^-\right)\right\} \right. \nn\\
& \left.+ \sinh(\Delta\Gamma t/2) \left\{\cos \zeta^- \left(\sin \phi^+ -
\frac{1}{2}{\rm Re}\delta \sin \phi^-\right) - \frac{1}{2}{\rm Re}\delta \sin \zeta^- \cos \phi^-\right\} \right.\nn\\
& \left.+\frac{1}{2} \cos(\Delta Mt) {\rm Im}\delta \cos\zeta^- \cos\phi^+ -\frac{1}{2} \sin(\Delta Mt)
 {\rm Im}\delta \sin\zeta^-\cos\phi^-\right]\,.
\end{align}

If there are multiple decay channels, one can generalize the above expression, by replacing
$\zeta^-,\phi^-,\phi^+$ with $\zeta_{l,m}^-$ etc., $|A_i||A_\perp|$ with $|A_i^m||A_\perp^l|$ and then taking
a summation over $l$ and $m$.

Then the two ``true" CP violating time-integrated triple product asymmetries
($i=0, \parallel$) for untagged decays are proportional to
\begin{align}
\Gamma \int_0^\infty {\rm Im}[A_\perp(t) A^*_i(t) + \bar A_\perp(t) \bar A^*_i(t)] &= \sum_{l,m} 2 |A^m_i||A^l_\perp| \times\nn\\
&\left[\left\{\cos\zeta_{l,m}^- \sin\phi^-_{l,m} - \frac{1}{4}{\rm Im}\delta \cos\phi^+_{l,m}
\left(1 + \sin\zeta_{l,m}^-\right)\right\}
\right. \nn\\
& \left.+ \left(\frac{\Delta \Gamma}{2 \Gamma}\right) \left\{\cos\zeta_{l,m}^- \left(\sin\phi^+_{l,m} -
\frac{1}{2}{\rm Re}\delta \sin \phi^-_{l,m}\right) -
\frac{1}{2}{\rm Re}\delta \sin\zeta_{l,m}^- \cos\phi^-_{l,m}\right\} \right.\nn\\
& \left.+\frac{1}{2} \left(\frac{1}{1 + \left(\frac{\Delta M}{\Gamma}\right)^2}\right)
{\rm Im}\delta \cos\zeta_{l,m}^- \cos\phi^+_{l,m}\right.\nn\\
& \left. -\frac{1}{2} \left(\frac{\frac{\Delta M}{\Gamma}}{1 + \left(\frac{\Delta M}{\Gamma}\right)^2}\right)
{\rm Im}\delta \sin\zeta_{l,m}^- \cos\phi^-_{l,m}\right]\,.
\end{align}
In the limit $\Delta M/\Gamma \ll 1$, one can neglect the last term and simplify the expression considerably.

We also note that even in the case $\zeta_{l,m}^- = \phi^-_{l,m} = 0$, {\em i.e.} when
all strong and weak phase differences cancel out individually, there is a nonzero TP asymmetry that
gives a clean measurement of ${\rm Im}\delta$:
\begin{align}
\Gamma \int_0^\infty {\rm Im}[A_\perp(t) A^*_i(t) + \bar A_\perp(t) \bar A^*_i(t)] \approx
\sum_{l,m} \frac12 |A^m_i||A^l_\perp| {\rm Im}\delta \cos\phi^+_{l,m}\,,
\end{align}
where we have used $\Delta M/\Gamma \approx 0$ and neglected the subleading $\Delta\Gamma/\Gamma$ terms.

\section{$B_s\to\phi\phi$ at LHCb}

The LHCb collaboration has recently measured the transversity amplitudes for the decay $B_s\to\phi\phi$
\cite{lhcb-cpt}, which is a pure penguin process and hence dominated by a single amplitude in the SM.
Thus, for all $l,m$, $A^l_i = A^m_i$ (for $i = 0,
\parallel, \perp$) The analysis also assumes that the weak phases of the three polarization
amplitudes are all equal; thus, all $\phi^{i-}_{l,m}$ (for $i = 0, \parallel$) in our
notation become zero. The correspondence between our notation and that of Ref.~\cite{lhcb-cpt} is as
follows:
\begin{align}
\label{transpose}
\mathcal A^{(2){\rm untagged}}_T &\to A_U\,,\ \ \
\mathcal A^{(1){\rm untagged}}_T \to A_V \nn \\
\left(\zeta_{\perp}- \zeta_{\parallel}\right) &\to \delta_1\,,\ \
\left(\zeta_{\perp}- \zeta_0 \right) \to \delta_2\,,\ \
\left(\zeta_{\parallel}- \zeta_0\right) \to \delta_{\parallel} \equiv \left(\delta_2- \delta_1\right)\,.
\end{align}

With the standard normalization of the transversity amplitudes, {\em viz.}
$|A_0|^2 + |A_\perp|^2 + |A_\parallel|^2 = |\bar A_0|^2 +
 |\bar A_\perp|^2 + |\bar A_\parallel|^2 = 1$, Eq.~(\ref{adecay}) becomes

\begin{align}
\label{finalexp}
\nn A_V &= - \frac{2 \sqrt2}{\pi} |A_\perp||A_0|\left[ - \sin(\delta_2)
\left({\rm Re}(f)\right)+\left\{ \cos(\delta_2) \sin(2 \phi_s)+ \sin(\delta_2)
\left({\rm Re}(f) \cos(2 \phi_s) + {\rm Im}(f) \sin(2 \phi_s)\right)\right\}(\Delta\Gamma/2\Gamma)\right]\\
 \nn &+ {\cal O}[(\Delta\Gamma/2\Gamma)^2]\\
\nn A_U &= - \frac{4}{\pi} |A_\perp||A_\parallel|\left[ - \sin(\delta_1) {\rm Re}(f)+\left\{ \cos(\delta_1) \sin(2 \phi_s)+
\sin(\delta_1) \left({\rm Re}(f) \cos(2 \phi_s) + {\rm Im}(f) \sin(2 \phi_s)\right)\right\}(\Delta\Gamma/2\Gamma)\right]\\
 &+ {\cal O}[(\Delta\Gamma/2\Gamma)^2]\,.
\end{align}

We will use the following numbers from Ref.~\cite{lhcb-cpt}:
\begin{align}
\label{expt}
|A_0|^2 &= 0.365\pm 0.022 (\text{stat})\pm 0.012(\text{syst}), \nn\\
|A_{\perp}|^2 &= 0.291\pm 0.024 (\text{stat})\pm 0.010(\text{syst}), \nn\\
|A_{\parallel}|^2 &= 0.344\pm 0.024 (\text{stat})\pm 0.014(\text{syst}), \nn\\
\cos(\delta_{\parallel}) &= -0.844\pm 0.068(\text{stat})\pm 0.029 (\text{syst}), \nn\\
A_U &= -0.055\pm 0.036 (\text{stat})\pm 0.018 (\text{syst}) \nn\\
A_V &= 0.010\pm 0.036 (\text{stat})\pm 0.018 (\text{syst})\,.
\end{align}

\begin{figure}[htbp]
       \includegraphics[scale=0.4]{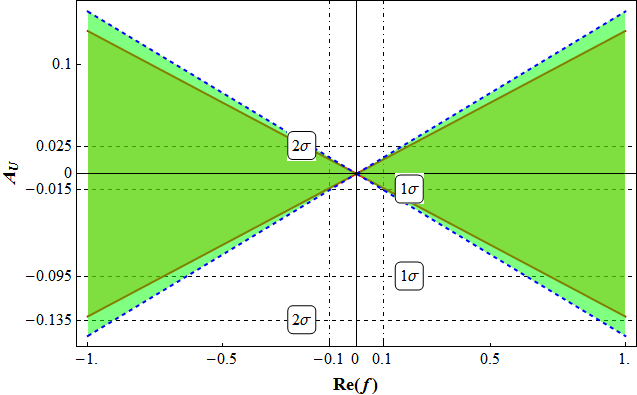}~~
       \includegraphics[scale=0.4]{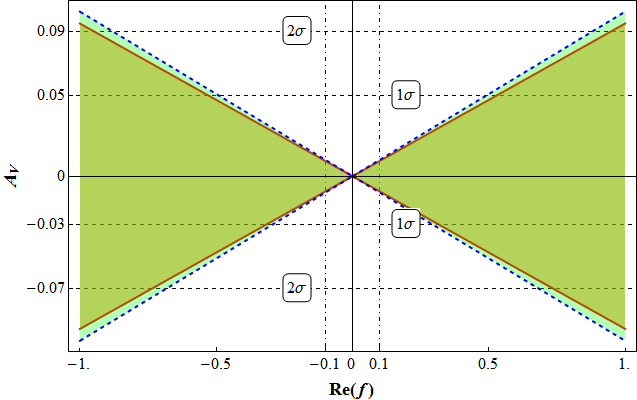}
\caption{Left panel: Allowed values of $A_U$ for $-1 \leq {\rm Re}(f) \leq 1$. The inner wedge is
for the input parameters varied in their $1\sigma$ ranges, the outer wedge is for $2\sigma$ variation.
Also shown are the $1\sigma$ and $2\sigma$ experimental bands for $A_U$, and the allowed region for a
smaller range of ${\rm Re}(f)$, namely, $|{\rm Re}(f)|\leq 0.1$. Right panel: Same plot for $A_V$.}
\label{fig:auav}
 \end{figure}

\begin{figure}[htbp]
       \includegraphics[scale=0.7]{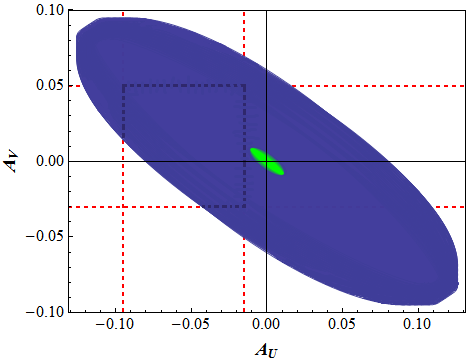}~~
       \includegraphics[scale=0.7]{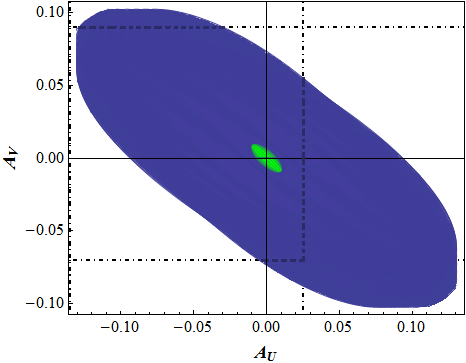}
\caption{Left panel: Allowed region in the $A_U$-$A_V$ plane when all the input parameters are
varied over their $1\sigma$ ranges. The outer ellipse is for $-1 \leq {\rm Re}(f) \leq 1$ and the
inner green ellipse is for $-0.1 \leq {\rm Re}(f) \leq 0.1$. The $1\sigma$ bands for $A_U$ and $A_V$
are shown as dashed lines. Right panel: The same plot when the input parameters are varied over
$2\sigma$; also, the $2\sigma$ bands are shown. The left edge of the $A_U$ band coincides with the
left edge of the plot.}
\label{fig:realf}
 \end{figure}

For our analysis, we use Eqs.~(\ref{transpose}), (\ref{finalexp}) and (\ref{expt}), and keep
terms only up to the first order in $\Delta\Gamma/\Gamma$. Even for the $B_s$ system, this is a
good approximation. All $\phi^{i+}_{l,m}$s in Eq.~(\ref{int}) (for $i = 0, \parallel$)
are now equal to $2 \phi_s$,  where $\phi_s$ is
the weak CP violating phase which is the same for the three polarization amplitudes, and very small
in the SM ($\phi_s \sim 0.02$ \cite{Beneke:2006hg,Bartsch:2008ps} based on QCD factorization)
\footnote{This should not be confused with the phase $\phi_s$ relevant for $\bsbsbar$ mixing and defined
as $\phi_s = {\rm arg}(-M_{12}/\Gamma_{12})$.}.
Even if there is some new physics making $\phi_s$ large,
the effects will be suppressed by $\Delta\Gamma/\Gamma$, so we do not expect much sensitivity
on the precise value of $\phi_s$. One may note that this phase has recently been measured by the LHCb 
collaboration \cite{1303.7125} to be between $-2.46$ and $-0.76$ rad with $68\%$ confidence 
level, which is not exactly in total conformity with the SM prediction. 

As is evident from Eq.~(\ref{finalexp}), if we neglect higher order terms in $\Delta\Gamma/\Gamma$,
both $A_U$ and $A_V$ are zero in the SM; thus, any definite nonzero value for these observables would point to
the presence of some NP. Considering CPT violation as the source of NP, one sees that there is a definite deviation
from zero even at the zero-th order of $\Delta\Gamma/\Gamma$; unfortunately, the shift depends only on
${\rm Re}(f)$, as ${\rm Im}(f)$ comes as a coefficient of $\sin(2\phi_s)$ in the subleading order.
Fig.~\ref{fig:auav} shows the allowed ranges for $A_U$ and $A_V$ when the input parameters are
varied over their experimental ranges. We have varied the three transversity amplitudes over their
allowed ranges keeping the normalization to unity fixed, and also varied the strong phase differences
$\delta_1$ and $\delta_2$ over the entire range of $[0:2\pi]$ keeping the constraint on $\cos(\delta_\parallel)$.
This gives a bound on $A_U$ and $A_V$, although this is quite weak at present (however, note that
if we take the $1\sigma$ region on $A_U$ seriously, small values of ${\rm Re}(f)$ are ruled out, as is the
SM). The allowed region will shrink considerably with more data.

In Fig.~\ref{fig:realf} we show the allowed region in the $A_U$-$A_V$ plane for large and small values
of ${\rm Re}(f)$, varying all other input parameters as above. Again, with more data, the elliptic figures are
bound to shrink, as well as the horizontal and vertical bands, constraining CPT violation. If finally the
intersection of the bands settle outside the ellipses, that will rule out CPT violation in this channel
at least, but that will also rule out the pure-SM explanation and call for some other NP.

\section{Conclusions}

The role of TP asymmetries as a probe of CP violation crucially hinges on the CPT theorem which relates
a possible T violating observable to a CP violating one. If CPT is not conserved, there is no such relationship,
and observables that are not supposed to show any TP asymmetries in the SM might do so. For example, if
CPT violation is present in one or more decay amplitudes, there will be a nonzero TP asymmetry even if the
weak phases of all the amplitudes are equal. The same trend persists in the time-dependence of the TP
asymmetries.

One might trade the $s$, $p$, and $d$-wave amplitudes with the transversity amplitudes, which are
directly accessible to the experiments. Some of the interference terms between these anplitudes are
CP violating only if the corresponding weak phases are different; in the presence of CPT violation,
we again observe that a nonzero signal can be observed even if all the weak phases are equal. The
observables $A_U$ and $A_V$, as measured by LHCb, are supposed to be zero in the SM
for channels like $B_s\to\phi\phi$. We show how one gets nonzero and possibly large
values for these observables
with CPT violation; a more canonical NP that contributes only to the $\bsbsbar$ mixing and hence
modifies the weak CP violating phase $\phi_s$ in the decay can hardly generate such large values as
all $\phi_s$-dependent terms are suppressed by $\Delta\Gamma/\Gamma$. The other side of the coin is
that with more data, one can successfully constrain the parameter space for the CPT violating parameters.

\acknowledgments

SKP was supported by a fellowship from UGC, Government of India.
AK was supported by CSIR (project no.\ 03(1135)/9/EMR-II),
and also by the DRS programme of the UGC, all under the Government
of India.


\appendix
\section{Factorization}

Following Ref.~\cite{datta}, we briefly describe the main results of naive factorization.
The prediction of naive factorization, that most TP asymmetries with ground state vector mesons are expected to be small
in the SM, will necessarily hold in PQCD or QCD factorization too.


The starting point for factorization is the SM effective hamiltonian
for $B$ decays \cite{BuraseffH}:
\bea
H_{eff}^q = {G_F \over \protect \sqrt{2}}
[V_{fb}V^*_{fq}(c_1O_{1f}^q + c_2 O_{2f}^q) - \sum_{i=3}^{10}(V_{ub}V^*_{uq} c_i^u
+V_{cb}V^*_{cq} c_i^c +V_{tb}V^*_{tq} c_i^t) O_i^q] + h.c.,
\label{Heff}
\eea
where the superscript $u$, $c$, $t$ indicates the internal quark, $f$
can be the $u$ or $c$ quark, and $q$ can be either a $d$ or $s$ quark.

Within factorization, the amplitude for $B\to V_1 V_2$ can be written
as
\be
{\cal A}(B \to V_1 V_2) = \sum_{{\cal O},{\cal O}'} \left\{ \bra{V_1}
{\cal O} \ket{0} \bra{V_2} {\cal O}' \ket{B} + \bra{V_2} {\cal O}
\ket{0} \bra{V_1} {\cal O}' \ket{B} \right\} ~,
\label{2amps}
\ee
where ${\cal O}$ and ${\cal O}'$ are some relevant four-fermion operators.
The first amplitude, $\bra{V_1}{\cal O} \ket{0}$, is proportional to the polarization
vector of $V_1$, namelt, $\varepsilon_1^\ast$. The second amplitude,
$\bra{V_2} {\cal O}' \ket{B}$, can be written
in terms of the usual vector and axial-vector form factors. Thus, the first term
of Eq.~(\ref{2amps}) is given by
\bea
\sum_{{\cal O},{\cal O}'} \bra{V_1} {\cal O} \ket{0} \bra{V_2} {\cal
O}' \ket{B}& =& -(m_B+m_2)m_1g_{V_1}XA_1^{(2)}(m_1^2)
 \varepsilon_1^* \cdot \varepsilon_2^* + 2\frac{m_1}{m_B+m_2}g_{V_1}X A_2^{(2)}(m_1^2) \varepsilon_2^*
\cdot p \varepsilon_1^* \cdot p \nn\\
&& ~ -i \frac{m_1}{ (m_B+m_2)}g_{V_1}X V^{(2)}(m_1^2)\epsilon_{\mu
\nu\rho\sigma} p^\mu q^\nu\varepsilon_1^{*\rho}
\varepsilon_2^{*\sigma}\,,
\label{term1}
\eea
All phase information is contained within the factor $X$, which is common to all
the three independent amplitudes. Thus, these quantities must have the
same phase.

A similar treatment for the second term in Eq.~(\ref{2amps}) gives
\bea
\sum_{{\cal O},{\cal O}'} \bra{V_2} {\cal O} \ket{0} \bra{V_1} {\cal
O}' \ket{B}& =& -(m_B+m_1)m_2g_{V_2}YA_1^{(1)}(m_2^2) \varepsilon_1^*
\cdot \varepsilon_2^* + 2\frac{m_2}{m_B+m_1}g_{V_2}Y A_2^{(1)}(m_2^2) \varepsilon_2^*
\cdot p \varepsilon_1^* \cdot p \nn\\
&& ~ -i \frac{m_2}{ (m_B+m_1)}g_{V_2}Y V^{(1)}(m_2^2)\epsilon_{\mu
\nu\rho\sigma} p^\mu q^\nu\varepsilon_1^{*\rho}
\varepsilon_2^{*\sigma}\,,
\label{term2}
\eea
where the phase informations are contained in the common factor $Y$, which need not be
the same as $X$.

We can now express the quantities $a$, $b$ and $c$ of
Eq.~(\ref{abcdefs}) as follows:
\bea
a &=& -m_1g_{V_1}(m_B+m_2) A_1^{(2)}(m_1^2) X - m_2g_{V_2}(m_B+m_1)
A_1^{(1)}(m_2^2) Y\nn\\
b &=& 2m_1g_{V_1}{m_B \over (m_B+m_2)}m_B A_2^{(2)}(m_1^2) X +
2m_2g_{V_2}{m_B \over (m_B+m_1)}m_B A_2^{(1)}(m_2^2) Y\nn\\
c &=& -m_1g_{V_1}{m_B \over (m_B+m_2)}m_B V^{(2)}(m_1^2) X
-m_2g_{V_2}{m_B \over (m_B+m_1)}m_B V^{(1)}(m_2^2) Y ~.
\label{abc}
\eea

Thus, nonzero TP asymmetries are generated from
${\rm Im}(a c^*)$ or ${\rm Im}(b c^*)$
if and only if both $X$ and $Y$ are present with different phase. Thus, if $V_1=V_2$, there cannot be
any TP asymmetry in the SM.

\section{CPT Violation in mixing}\label{cptmixform}

This closely follows Ref.~\cite{patra} with a coupe of typographical errors corrected.
Consider the $2\times 2$ Hamiltonian matrix with an explicit CPT violating term $\delta$.
Let us define,
\be
 \eta_1 = \frac{q_1}{p_1} = \left(y + \frac{\delta}{2}\right) \alpha\,;
\ \ \
\eta_2 = \frac{q_2}{p_2} = \left(y - \frac{\delta}{2}\right) \alpha\,;
\ \ \
\omega = \frac{\eta_1}{\eta_2}\,,
\ee
and
\begin{align}
\nonumber f_-(t) & =
\frac{1}{(1 + \omega)} \left(e^{-i\lambda_1 t} - e^{-i\lambda_2 t}\right)\,, \\
\nonumber f_+(t) & =\frac{1}{(1 + \omega)} \left(e^{-i\lambda_1 t} + \omega e^{-i\lambda_2 t}\right)\,, \\
 \bar f_+(t) & =\frac{1}{(1 + \omega)} \left(\omega e^{-i\lambda_1 t} + e^{-i\lambda_2 t}\right)\,.
\end{align}
Thus,
\begin{align}
 \left|f_-(t)\right|^2 & =
\frac{2 e^{-\Gamma t}}{|1+\omega|^2} \left[\cosh\left(\frac{\Delta\Gamma t}{2}
\right) - \cos\left(\Delta{M} t\right)\right]\, \approx \frac{e^{-\Gamma t}
\left(1 - {\rm Re}\delta\right)}{2} \left[\cosh\left(\frac{\Delta\Gamma t}{2}
\right) - \cos\left(\Delta{M} t\right)\right]\,,\nn\\
 \left|f_+(t)\right|^2 & =
\frac{e^{-\Gamma t}}{|1+\omega|^2} \Bigg[ \cosh\left(\frac{\Delta \Gamma t}{2}
\right) (1 + |\omega|^2) + \sinh\left(\frac{\Delta \Gamma t}{2}
\right) (1 - |\omega|^2) + 2 {\rm Re}(\omega) \cos\left(\Delta M t\right) - 2 {\rm Im}(\omega)
\sin\left(\Delta M t\right)\Bigg]\,, \nonumber \\
& \approx \frac{e^{-\Gamma t}}{2} \Bigg[ \cosh\left(\frac{\Delta \Gamma t}{2}
\right) - \sinh\left(\frac{\Delta \Gamma t}{2}
\right) {\rm Re}\delta + \cos\left(\Delta M t\right) - {\rm Im}\delta
\sin\left(\Delta M t\right)\Bigg]\,,\nn\\
\left|\bar f_+(t)\right|^2 & =
 \frac{e^{-\Gamma t}}{|1+\omega|^2} \Bigg[ \cosh\left(\frac{\Delta \Gamma t}{2}
\right)(1+|\omega|^2) - \sinh\left(\frac{\Delta \Gamma t
}{2}\right)(1-|\omega|^2) + 2{\rm Re}(\omega) \cos\left(\Delta M t\right) + 2 {\rm Im}(\omega)
\sin\left(\Delta M t\right)\Bigg]\,,\nonumber \\
& \approx \frac{e^{-\Gamma t}}{2} \Bigg[ \cosh\left(\frac{\Delta \Gamma t}{2}
\right) + \sinh\left(\frac{\Delta \Gamma t}{2}
\right) {\rm Re}\delta + \cos\left(\Delta M t\right) + {\rm Im}\delta
\sin\left(\Delta M t\right)\Bigg]\,, \nonumber \\
 f_+^*(t) f_- (t) & = \frac{e^{-\Gamma t}}{|1+\omega|^2}
\Bigg[ \cosh\left(\frac{\Delta \Gamma t}{2}\right)(1-\omega^{\ast}) +
\sinh\left(\frac{\Delta \Gamma t }{2}\right)(1+\omega^{\ast}) + \cos\left(\Delta M t\right) (-1 + \omega^{\ast}) - i \sin
\left(\Delta M t\right)(1+ \omega^{\ast}) \Bigg]\,, \nonumber \\
& \approx \frac{e^{-\Gamma t}}{4}
\Bigg[ \cosh\left(\frac{\Delta \Gamma t}{2}\right)(- {\rm Re}\delta + i {\rm Im}\delta) +
\sinh\left(\frac{\Delta \Gamma t }{2}\right)(2 - {\rm Re}\delta - i {\rm Im}\delta) \nn\\
& + \cos\left(\Delta M t\right) ({\rm Re}\delta - i {\rm Im}\delta) - \sin
\left(\Delta M t\right)\left({\rm Im}\delta + i (2 - {\rm Re}\delta)\right) \Bigg]\,, \nonumber \\
\bar{f_+}(t) f^{\ast}_{-} (t) & =
\frac{e^{-\Gamma t}}{|1+\omega|^2} \Bigg[ \cosh\left(\frac{\Delta \Gamma t}{2}
\right)(\omega-1) +
\sinh\left(\frac{\Delta \Gamma t }{2}\right)(1+\omega) + \cos\left(\Delta M t\right) (1 - \omega) +
i \sin\left(\Delta M t\right)(1+ \omega) \Bigg]\, \nn\\
& \approx
\frac{e^{-\Gamma t}}{4} \Bigg[ \cosh\left(\frac{\Delta \Gamma t}{2}
\right)({\rm Re}\delta + i {\rm Im}\delta) +
\sinh\left(\frac{\Delta \Gamma t }{2}\right)(2 - {\rm Re}\delta + i {\rm Im}\delta) \nn\\
& - \cos\left(\Delta M t\right) ({\rm Re}\delta + i {\rm Im}\delta) +
i \sin\left(\Delta M t\right)\left(-{\rm Im}\delta + i (2- {\rm Re}\delta)\right) \Bigg]\,,
\end{align}
Where we take, $y\approx 1\,, \eta_{1(2)} \approx \left(1 +(-) \frac{\delta}{2}\right)\,,
\omega \approx \left(1 + \delta\right)\,, |\omega|^2 \approx \left(1 + 2 {\rm Re}\delta\right)\,,
|1 + \omega|^{-2} \approx \frac14(1 - {\rm Re}\delta)\,, |\eta_{1(2)}|^2 \approx
\left(1 +(-) {\rm Re}\delta\right)$.

This gives,
\begin{align}
A^*_i(t)A_k(t) &= \left[f^*_+A^*_i + \eta_1^* f^*_-\bar A^*_i\right] \left[f_+ A_k + \eta_1 f_- \bar A_k\right]
\nonumber\\
&=A^*_iA_k \left[|f_+|^2 + \eta_1 (\bar A_k/A_k) f^*_+f_- \right]
+ \bar A^*_i\bar A_k \left[ |\eta_1|^2 |f_-|^2 + \eta_1^* (A_k/\bar A_k) f_+f^*_- \right]
\nonumber \\
&= \frac{e^{-\Gamma t}}{2}\left[ A^*_i A_k \left\{ \cosh(\Delta\Gamma t/2) + \cos(\Delta mt) - {\rm Re} \delta \sinh(\Delta\Gamma t/2) - {\rm Im} \delta \sin(\Delta mt)\right\} \right. \nn \\
&\left.+ \frac{\eta_k e^{-2i\phi_k}}{2} A^*_i A_k \left\{2 \sinh(\Delta\Gamma t/2) - 2 i\sin(\Delta Mt) + \left(- {\rm Re}\delta +
i {\rm Im}\delta\right) \cosh(\Delta\Gamma t/2) + \left({\rm Re}\delta - i {\rm Im}\delta\right) \cos(\Delta Mt)\right\} \right.
\nonumber\\
&~~~~~~~~~~\left.+\bar A^*_i\bar A_k \left\{ \cosh(\Delta\Gamma t/2) - \cos(\Delta Mt)\right\}\right. \nn\\
&\left.+ \frac{\eta_ke^{2i\phi_k}}{2} \bar A^*_i\bar A_k \left\{2 \sinh(\Delta\Gamma t/2) + 2 i\sin(\Delta Mt) +
\cosh(\Delta\Gamma t/2) \left(-{\rm Re}\delta - i {\rm Im}\delta\right)+ \left({\rm Re}\delta + i {\rm Im}\delta\right)
\cos(\Delta mt)\right\} \right]\,,
\nonumber
\end{align}
\begin{align}
\label{aitimecpt}
\bar A^*_i(t) \bar A_k(t) &= \left[\frac{f^*_-}{\eta_2^*} A^*_i + \bar f^*_+ \bar A^*_i\right]
\left[\bar f_+ \bar A_k + \frac{f_-}{\eta_2} A_k \right]
\nonumber\\
&=A^*_iA_k\left[\frac{|f_-|^2}{|\eta_2|^2} + (\bar A_k/A_k) \frac{\bar f_+f^*_-}{\eta_2^*}\right]
+ \bar A^*_i\bar A_k\left[|\bar f_+|^2 + (A_k/\bar A_k)\frac{f_- \bar f^*_+}{\eta_2}\right]
\nonumber \\
&= \frac{e^{-\Gamma t}}{2}\left[ A^*_iA_k \left\{ \cosh(\Delta\Gamma t/2) - \cos(\Delta Mt) \right\}\right. \nn \\
&\left. + \frac{\eta_k e^{-2i\phi_k}}{2} A^*_iA_k \left\{ 2 \sinh(\Delta\Gamma t/2) + 2 i \sin(\Delta Mt) +
\left({\rm Re} \delta + i {\rm Im} \delta\right) \cosh(\Delta\Gamma t/2) - \left({\rm Re} \delta + i {\rm Im} \delta\right)
\cos(\Delta Mt) \right\} \right.
\nonumber\\
&~~~~~~~~~~\left.+ \bar A^*_i\bar A_k \left\{ \cosh(\Delta\Gamma t/2) + \cos(\Delta Mt) + {\rm Re}\delta \sinh(\Delta\Gamma t/2)
+ {\rm Im}\delta \sin(\Delta Mt) \right\} \right. \nn \\
&\left.+ \frac{\eta_k e^{2i\phi_k}}{2} \bar A^*_i\bar A_k  \left\{ 2 \sinh(\Delta\Gamma t/2) -
2 i\sin(\Delta Mt) + \left({\rm Re} \delta - i {\rm Im} \delta\right)
\cosh(\Delta\Gamma t/2) - \left({\rm Re} \delta - i {\rm Im} \delta\right) \cos(\Delta Mt) \right\} \right]\;.
\end{align}


\end{document}